\documentclass[11pt]{article}

\usepackage{amsmath}
\usepackage{amssymb}
\usepackage{bbm}
\usepackage{cancel}
\usepackage{mathtools}
\usepackage{slashed}
\usepackage[numbers,sort&compress]{natbib}

\usepackage[letterpaper,margin=1in,bottom=1in]{geometry}
\usepackage{float} 
\usepackage{parskip} 
\usepackage{tabulary} 
\usepackage{color} 
\usepackage{soul} 
\usepackage{subfigure}
\usepackage{graphicx}
\usepackage[section]{placeins} 

\def\gco{gluino-neutralino~coannihilation}
\def\GCO{Gluino-Neutralino~Coannihilation}
\def\lhc2{LHC~RUN~II}

\def\nsu{non-universal~supergravity~model}

\usepackage{cleveref}

\newcommand{\code}[1]{\texttt{#1}}

\def \na{\widetilde{\chi}^{0}}

\def \g{\tilde{g}}

\def\.4{\vspace{-.5cm}}

\def\beq{\begin{equation}}
\def\be{\begin{equation}}
\def\beqn{\begin{eqnarray}}
\def\ee{\end{equation}}
\def\eeq{\end{equation}}
\def\eeqn{\end{eqnarray}}

\author{
Pran Nath\footnote{Email:p.nath@neu.edu}~\,and
Andrew B. Spisak\footnote{Email:a.spisak@neu.edu}
\\~\\
Department of Physics, Northeastern University,
Boston, MA 02115-5000, USA
}

\title{Gluino Coannihilation and Observability of Gluinos at \lhc2
}

\begin{document}
\maketitle
\date

\textbf{Abstract: }
The observability of a gluino at \lhc2 is analyzed for the case where the gluino lies in the \gco~region and  the mass gap between the gluino and the neutralino is small. The analysis is carried out under the Higgs boson mass constraint and the constraint of  dark matter relic density consistent with the WMAP and Planck experiment. It is shown that in this case  a gluino with mass much smaller than the current lower limit of $\sim 1500$ GeV as given by \lhc2  at 3.2 fb$^{-1}$ of integrated luminosity would have escaped detection. The analysis is done using the signal regions used by the ATLAS Collaboration where an optimization of signal regions was carried out to determine the best regions for
gluino discovery  in the \gco~region.
It is shown that under the Higgs boson mass constraint and the  relic density constraint, a gluino  mass  of $\sim 700$ GeV would require $14$ fb$^{-1}$ of integrated luminosity for discovery and a gluino of mass $\sim 1250$ {\rm GeV} would require 3400 fb$^{-1}$ of integrated luminosity for discovery at \lhc2. An analysis of dark matter for this case is also given. It is found that for the range of gluino masses considered, the neutralino mass lies within less than 100 GeV of the gluino mass.  Thus a measurement  of the gluino mass in the \gco ~region will provide a 
determination of the neutralino mass. In this region the neutralino is 
dominantly  a gaugino  and the  spin-independent proton-neutralino cross section is small but much larger than
the neutrino floor  lying in the range  $\sim (1-10)\times 10^{-47}$ cm$^{2}$.  Thus a  significant part of the 
parameter space of the model will lie within the reach of the next generation LUX-ZEPLIN dark matter experiment. 

\section{Introduction}\label{sec:intro}

The fact that sparticles  have not been observed so far appears to imply that the scale of weak scale supersymmetry (SUSY) is higher than previously thought, lying in the several TeV region. This reasoning receives support from the observation that the Higgs boson~\cite{Englert:1964et, Higgs:1964pj, Guralnik:1964eu} mass measurements give a mass of $\sim 126$ GeV,~\cite{Chatrchyan:2012ufa, Aad:2012tfa} while in supersymmetry the tree-level Higgs boson mass must lie below the mass of the Z boson. Thus the extra mass must arise from loop corrections, which then requires that the SUSY scale be in the several-TeV region~\cite{Akula:2011aa,susy-higgs}.
However, softly broken supersymmetry has more than one mass scale and a look at the loop correction indicates that the SUSY scale that enters is the third generation mass. Specifically, the gaugino masses are largely unconstrained and their scale could be much lower, allowing their early discovery. On the other hand, LHC data also appears to put significantly high limits, i.e. above 1.5 TeV, for the gluino mass~\cite{ATLAS:2015}.

In this work we show that in the \gco~region, gluino masses lower than 1 TeV would have escaped detection at LHC RUN I and also in the analyses at \lhc2~based on the accumulated integrated 
luminosity of 3.2 fb$^{-1}$.~\cite{ATLAS:2015}. Specifically we carry out an analysis in the framework of a high-scale supergravity grand unified model~\cite{msugra} with radiative breaking of the electroweak symmetry (for a reviews see~\cite{msugra-review}).
The supergravity unified models admit a large landscape for sparticle masses \cite{landscape} which allows for low mass gluinos \cite{landscape,Feldman:2009zc,Baer:1998pg,Profumo} and specifically sparticle landscapes where the gluino is the next-to-lightest supersymmetric particle (NLSP). The largeness of the sparticle landscape arises due the fact that one may have non-universal 
supergravity models~\cite{Ellis:1985jn,NU,nonuni2}.

We investigate the sparticle spectrum in the \gco~region under the Higgs boson mass constraints and the relic density constraint. In these models the universal scalar mass is taken to be high, lying in the several TeV region, but {{with low-lying gluino masses}. Models we consider allow for radiative breaking of the electroweak symmetry and lie on the hyperbolic branch,~\cite{hb} where large scalar masses can arise with low fine tuning. A number of works have analyzed these SUGRA models~\cite{Baer:2012mv, Altunkaynak:2010we, Berger:2008cq, Conley:2010du, Roszkowski:2009ye, Fowlie:2012im, Kim:2013uxa}. 

Here  we consider the possibility that if the first and second generation GUT scale gaugino masses $m_1=m_2$ are high while the third-generation gaugino mass $m_3$ is low, there can exist light gluinos with masses close to the LSP mass that have evaded detection at the LHC so far, and may yet be discoverable at \lhc2. Coannihilation of a light gluino with mass near the LSP neutralino provides a mechanism for achieving a dark matter relic density consistent with experimental constraints from WMAP~\cite{Larson:2010gs} and Planck~\cite{Ade:2015xua}. The analysis shows that in the \gco~region, gluino masses in the range 700--1300 Gev exist consistent with Higgs boson mass constraint and relic density constraint and would have escaped detection so far, but would be accessible at \lhc2~with up to 3000 fb$^{-1}$ of integrated luminosity. A search for  signal regions for the discovery of the gluino in the \gco~region was carried out. 
{
For each model point seven signal regions as listed by the ATLAS Collaboration~\cite{ATLAS:2015} and displayed in~table \ref{tab:SR} were analyzed. In this table, $H_T$ is defined as the scalar sum of the transverse momentum of all jets, $M_{\text{eff}}(incl.)$ is defined as the scalar sum of $E^{\text{miss}}_T$ and the transverse momentum of all jets with $p_T>50$ GeV, while $M_{\text{eff}}(N_j)$ is the scalar sum of $E^{\text{miss}}_T$ and the transverse momentum of the first $N$ jets. Each signal region is defined by up to 12 different cuts  which are on jet $P_T$'s, $E_T^{\rm miss}$, etc., which are meant to reduce the background and enhance the signal.} 
It is found that over the entire gluino mass region analyzed, the optimum signal regions are $2jl$ and $2jm$ (using the nomenclature of ATLAS~\cite{ATLAS:2015}) where $2jl$ and $2jm$ are as defined in table \ref{tab:SR}.

The outline of the rest of the paper is as follows: In section~\ref{sec:spectra} we examine the parameter space of models with coannihilation of a light ($< 2$~TeV) gluino with the LSP under the constraints of the Higgs boson mass, WMAP and Planck relic density, and LHC RUN I exclusions on the sparticle mass spectra. In section~\ref{sec:lhc} we carry out a signature analysis of a subset of benchmark model points for \lhc2. 
{
For these we analyze the seven signal regions of table \ref{tab:SR} and determine the leading and the subleading  signal regions and the minimum integrated luminosity for LHC discovery for each of the model points analyzed. This allows us to determine the range of gluino masses which would have escaped detection thus far but would be accessible at \lhc2 with up to 3000 fb$^{-1}$ of integrated luminosity.} 
We also compare to the latest constraints from ATLAS~\cite{ATLAS:2015} using simplified models with 3.2 fb$^{-1}$ of integrated luminosity for the 13 TeV data. In section \ref{optimization} we  carryout a signal region optimization.
In section~\ref{sec:dm} we investigate the direct detection of dark matter for the benchmark points of section~\ref{sec:lhc}. Conclusions are given in section~\ref{sec:conclusion}. 

\section{Parameter Space for Gluino Coannihilation in Non-Universal SUGRA}\label{sec:spectra}
As mentioned in section~\ref{sec:intro}, \nsu s allow for the possibility of non-universalities in the gaugino masses. In this work we consider this possibility by assuming that the scalar masses are all universal at high scale but the gaugino masses are split, i.e.,  gaugino masses in the $U(1)$ and $SU(2)$ sectors are equal, while the $SU(3)_C$ gaugino mass is not. Specifically we assume the following parameter space for the model:
\begin{align}
    m_0,~A_0,~m_1=m_2\neq m_3,~\tan\beta,~\text{sgn}(\mu)
    \label{eq:params}
\end{align}
where $m_0$ is the universal scalar mass, $m_1=m_2$, {where $m_1, m_2$ are the gaugino masses for the 
$U(1)$ and $SU(2)$ sectors and $m_3$ is the $SU(3)_C$ gaugino mass, }
$A_0$ is the universal trilinear scalar coupling (all at the grand unification scale) and  $\tan\beta=\langle H_2 \rangle/\langle H_1\rangle$, where $H_2$ gives mass to the up quarks and $H_1$ gives mass to the down quarks and the leptons, and sgn$(\mu)$ is the sign of the Higgs mixing parameter which enters in the superpotential in the term $\mu H_1 H_2$. Using the above parameter space the sparticle spectrum is generated using \code{SoftSusy 3.5.1}~\cite{Allanach:2001kg} while the analysis of the relic density is done using \code{Micromegas 4.1.7}~\cite{Belanger:2004yn} and SLHA-format data files are processed using \code{PySLHA}~\cite{Buckley:2013jua}.

We wish to examine the region of \gco~which requires that the gluino be the next-to-lightest supersymmetric particle (NLSP). In high scale models this can be achieved by letting $m_3$ lie significantly lower than the common scale $m_1=m_2$, as shown in table \ref{tab1}. Thus in table \ref{tab1} we list several models which are chosen as illustrative examples. In generating the parameter set of table \ref{tab1} we have imposed the constraints that the Higgs boson mass obey $m_h= 125\pm 2$ GeV and
$\Omega h^2 =0.11\pm 0.013$ consistent with WMAP and Planck experiments. The ranges are chosen to take account of possible errors in theoretical computations given by codes. 

\Cref{tab1} shows that the gluino mass lies close to the neutralino mass with their mass difference lying in the range of $\sim(70-100)$ GeV and the gluino mass being around $1.1\times$ the neutralino mass over essentially the entire gluino mass range. It is also to be noted that the scale of $m_0$ is high, lying in the several TeV region, while the light stop mass lies in the range between 2-3 TeV. The largeness of the third generation scalar masses is what provides a large loop correction to the 
Higgs boson to lift its tree value to the desired experimentally-observed value. The reason for the selection of the illustrative models of table \ref{tab1} as noted earlier is the following: As discussed in section~\ref{sec:lhc} all of the models in table \ref{tab1} would have escaped detection at LHC RUN I but would be observable at \lhc2.

\section{Signal Analysis for \GCO~Models at LHC RUN-II}\label{sec:lhc}

\begin{figure}[t]
	\includegraphics[width=0.5\textwidth]{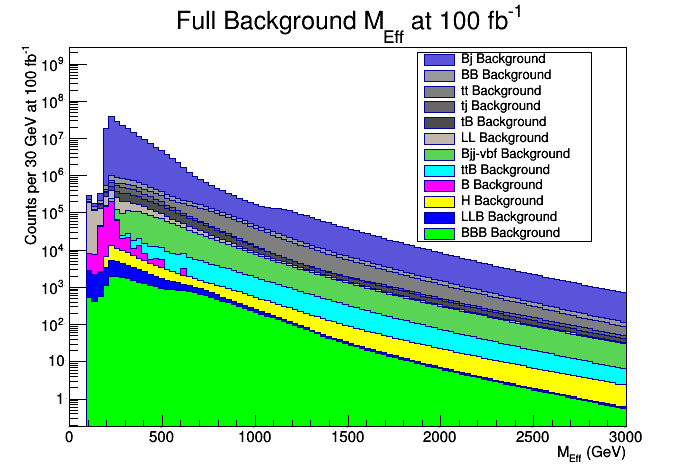}
    \includegraphics[width=0.5\textwidth]{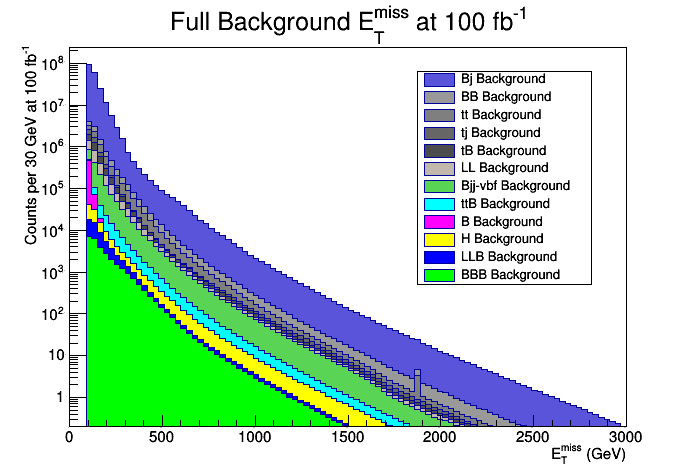}
    \caption{
    A display of Standard Model  SNOWMASS background after triggering cuts and a $E^{\rm miss}_T\geq 100\,{\rm GeV}$ pre-cut.  The background is broken into individual processes with the y-axis displaying the number of events per
    30 GeV  scaled to an integrated luminosity of 100 fb$^{-1}$ at \lhc2. The left panel displays 
    the distribution as a function of 
    $M_{\rm Eff}({\rm incl.})$ while the right panel displays the distribution as a function of 
    $E^{\rm miss}_T$. Individual background processes are colored according to eq.~(\ref{Snowlabels}).}
	\label{fig1}
\end{figure} 

After applying the Higgs boson and the relic density constraints to the \gco~region, a set of benchmark points was selected for a Monte Carlo analysis of LHC signal regions (see table \ref{tab1}). This analysis was performed with the \code{MadGraph 2.2.2}~\cite{Alwall:2014hca} software system. First, the Feynman diagrams were calculated for all possible decays of the form p p $\to$ SUSY SUSY, where ``SUSY'' can be any MSSM particle.  {The analysis is configured to include both ISR and FSR jets.} With the sparticle spectra of the benchmark points calculated by \code{SoftSUSY}, as well as the decay widths and branching ratios calculated by \code{SDECAY}~\cite{Muhlleitner:2004mka, Muhlleitner:2003vg} and \code{HDECAY}~\cite{Djouadi:1997yw,Spira:1996if} operating within \code{SUSY-HIT 1.5}~\cite{Djouadi:2006bz}, \code{MadEvent} was used to simulate 50,000 MSSM decay events for each benchmark point.  Hadronization of resultant particles is handled by \code{Pythia 6.4.28}~\cite{Sjostrand:2006za}, and ATLAS detector simulation and event reconstruction is performed by \code{Delphes 3.1.2}~\cite{deFavereau:2013fsa}.
A large set of search analyses were performed on the generated events for each benchmark point. The analyses used \code{ROOT 5.34.21}~\cite{Brun:1997} to implement search region details for the recently published 2-6 jet plus missing transverse energy signal search regions at 13~TeV~\cite{ATLAS:2015}, which look specifically for light squarks and gluons.

To allow comparison to the background, all of the signal region analyses were applied to pre-generated backgrounds published by the SNOWMASS group~\cite{Avetisyan:2013onh}. For each benchmark point, a calculated implied luminosity allowed direct comparison to the backgrounds.  Each individual background process from the SNOWMASS background set was scaled by its own implied luminosity and combined to determine a total background count for each signal region.  Fig.~\ref{fig1} displays
{distributions in two} key kinematic quantities, $M_{\rm eff}$ and $E_T^{\rm miss}$, for individual processes and their relative contribution to the overall background after minimal cuts for trigger simulation and a $E_T^{\rm miss} \geq 100\,{\rm GeV}$ pre-cut, but before any specific cuts for signal regions. Here and throughout,  $M_{\rm eff}$ is specifically  $M_{\rm eff}(incl.)$, defined as the scalar sum of $E_T^{\rm miss}$ and the $p_T$ of all jets with $p_T(j) \geq 50\,{\rm GeV}$. The various background samples are grouped according to the generated final state, with a collective notation given by
\begin{align} 
J = \left\lbrace u,\bar{u} ,d,\bar{d},s,\bar{s} ,c,\bar{c},b,\bar{b} \right\rbrace , \nonumber \\
L = \left\lbrace e^+,e^-,\mu^+,\mu^-,\tau^+,\tau^-,\nu_e,\nu_{\mu}, \nu_{\tau}\right\rbrace, \\
V = \left\lbrace W^+,W^-, Z , \gamma, h_0 \right\rbrace , \,
T = \left\lbrace t,\bar{t}\, \right\rbrace , \,  H = \left\lbrace h_0 \right\rbrace . \nonumber 
\label{Snowlabels} 
\end{align}

In general, events with gauge bosons and SM Higgs bosons in the final state are grouped into a single ``boson'' (B) category. Thus, for example,  the data set ``Bjj-vbf'' represents production via vector boson fusion of a gauge boson or a Higgs with at least two additional light-quark jets.

\subsection{LHC Production and Signal Definitions}
The
multi-jet search strategy  described in~\cite{ATLAS:2015} defines seven signal regions are defined with jet multiplicities ranging from 2~to~6 jets with cuts on the inclusive effective mass $M_{\rm eff}(incl.)$ varying from ``loose'' (low $M_{\rm eff}(incl.)$) to ``tight'' (high $M_{\rm eff}(incl.)$).  When examining the results for the gluino coannihilation benchmark points, $2jm$ and $2jl$ were found to be the dominant and subdominant signal regions for discovery.
\begin{table}[H]
	\centering
	\begin{tabulary}{\linewidth}{l|c|c|c|c|c|c|c}
    \hline\hline
	Requirement & \multicolumn{7}{c}{Value} \\
    \hline
    & 2jl & 2jm & 2jt & 4jt & 5j & 6jm & 6jt \\
    \cline{2-8}
    $E^{\text{miss}}_T\text{ (GeV)}>$ 
    & 200 & 200 & 200 & 200 & 200 & 200 & 200 \\
    $p_T(j_1)\text{ (GeV)}>$ 
    & 200 & 300 & 200 & 200 & 200 & 200 & 200 \\
    $p_T(j_2)\text{ (GeV)}>$ 
    & 200 & 50 & 200 & 100 & 100 & 100 & 100 \\
    $p_T(j_3)\text{ (GeV)}>$
    & $-$ & $-$ & $-$ & 100 & 100 & 100 & 100 \\
    $p_T(j_4)\text{ (GeV)}>$
    & $-$ & $-$ & $-$ & 100 & 100 & 100 & 100 \\
    $p_T(j_5)\text{ (GeV)}>$
    & $-$ & $-$ & $-$ & $-$ & 100 & 100 & 100 \\
    $p_T(j_6)\text{ (GeV)}>$
    & $-$ & $-$ & $-$ & $-$ & $-$ & 100 & 100 \\
    $\Delta\phi(\text{jet}_{1,2,(3),E^{\text{miss}}_T})_{\text{min}}>$ 
    & 0.8 & 0.4 & 0.8 & 0.4 & 0.4 & 0.4 & 0.4 \\
    $\Delta\phi(\text{jet}_{i>3,E^{\text{miss}}_T})_{\text{min}}>$ 
    & $-$ & $-$ & $-$ & 0.2 & 0.2 & 0.2 & 0.2 \\
    $E^{\text{miss}}_T/\sqrt{H_T}\text{ (GeV)}^{1/2}>$ 
    & 15 & 15 & 20 & $-$ & $-$ & $-$ & $-$ \\
    $E^{\text{miss}}_T/M_{\text{eff}}(N_j)>$ 
    & $-$ & $-$ & $-$ & 0.2 & 0.25 & 0.25 & 0.2 \\
    $M_{\text{eff}}(incl.)\text{ (GeV)}>$ 
    & 1200 & 1600 & 2000 & 2200 & 1600 & 1600 & 2000 \\
	\hline
	\end{tabulary}
	\caption{The selection criteria used for the signal regions in the nomenclature of Table 2 of the ATLAS analysis~\cite{ATLAS:2015}.}
	\label{tab:SR}
\end{table}

Using the techniques and signal regions described above, we analyzed each of the benchmark points of table \ref{tab1}
 to identify a signal region of minimum required luminosity for $5\sigma$ signal$/\sqrt{\text{background}}$ discovery of this point at the \lhc2.  These results can be directly compared to the results of~\cite{ATLAS:2015} for simplified models involving gluinos and neutralinos with decoupled squarks. {In that analysis it was found that for such a simplified model, using 3.2 fb$^{-1}$ of data on the same signal regions to detect the same gluino decay channel ($\tilde g \to q \bar q \tilde \chi_1^0$), a gluino mass lower limit of $\sim1500$~GeV is established for neutralino masses up to $\sim800$~GeV. For LSP masses above this, no exclusion is established at 3.2 fb$^{-1}$ for any gluino masses, demonstrating that the \gco~region remains viable for continued searching.}

\begin{table}[H]
\begin{center}
\begin{tabular}{l|ccccc|cccc}
\hline\hline
Model & Gluino & Neutralino & Stop & Higgs & $\Omega_{\text{LSP}}h^2$ & $m_0$ & $A_0$ & $m_1=m_2$ & $m_3$ \\
\hline
(i)   &$706$ &$634$ &$2124$&$123.8$&$0.122$&$5000$&$ -7000$&$1400$&$250$ \\
(ii)  &$836$ &$755$ &$3497$&$124.1$&$0.110$&$7000$&$ -8000$&$1650$&$300$ \\
(iii) &$955$ &$868$ &$2367$&$125.5$&$0.112$&$6000$&$ -9000$&$1900$&$350$ \\
(iv)  &$1057$&$975$ &$2754$&$123.2$&$0.102$&$5500$&$ -5500$&$2150$&$400$ \\
(v)   &$1129$&$1046$&$2910$&$123.5$&$0.101$&$5800$&$ -5800$&$2300$&$430$ \\
(vi)  &$1201$&$1107$&$2932$&$126.7$&$0.110$&$7500$&$-11500$&$2400$&$450$ \\
(vii) &$1252$&$1167$&$3459$&$124.1$&$0.101$&$6800$&$ -6800$&$2550$&$480$ \\
\hline
\end{tabular}\end{center}
\caption{A list of \gco~models which satisfy Higgs boson mass and relic density constraints for the case when $\tan\beta=10$ and sign($\mu$) is positive. All masses are in GeV. {Relic density constraints were determined by taking $\pm2.5~\times$ the WMAP7 error of $\pm0.0056$}}
\label{tab1}
\end{table}

\begin{figure}[H]
\begin{center}
	\includegraphics[width=0.7\textwidth]{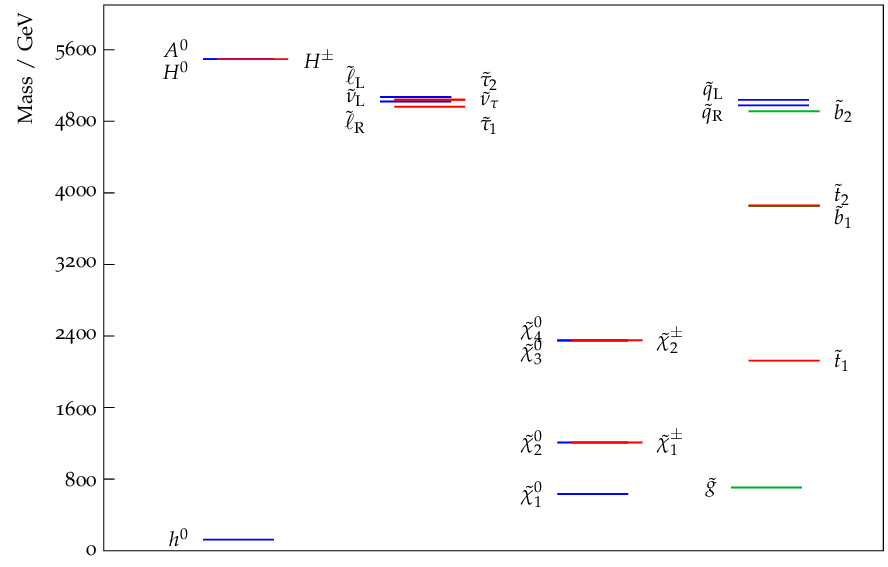}
\end{center}
    \caption{An exhibition of the sparticle mass hierarchy for the \gco~Model (i).}
    \label{fig2}
\end{figure} 

\begin{table}[H]
\begin{center}
\begin{tabulary}{0.85\textwidth}{l|CCCc}
\hline\hline
Model & SUSY Cross Section (pb)  & BR($\tilde g\to q\bar q \,\tilde \chi_1^0$), $q\in\{u,d\}$ & BR($\tilde g\to q\bar q \,\tilde \chi_1^0$), $q\in\{c,s,b\}$ & BR($\tilde g\to g\tilde \chi_1^0$) \\
\hline
(i)   & $2.8$   & 0.42 & 0.55 & 0.03\\
(ii)  & $0.95$  & 0.43 & 0.56 & 0.02\\
(iii) & $0.38$  & 0.43 & 0.56 & 0.01\\
(iv)  & $0.18$  & 0.38 & 0.49 & 0.13\\
(v)   & $0.11$  & 0.38 & 0.49 & 0.13\\
(vi)  & $0.071$ & 0.43 & 0.57 & 0.00\\
(vii) & $0.051$ & 0.39 & 0.51 & 0.10\\
\hline
\end{tabulary}\end{center}
\caption{SUSY production cross sections in picobarns and the gluino decay branching ratios into light quarks (u,d) and other non-top quarks (c,s,b) for \gco~Models (i)-(vii) of table \ref{tab1} at \lhc2. Gluinos are pair-produced by gluon fusion via the process $gg\to \tilde g \tilde g$ as well as by $q\bar q\to \tilde g \tilde g$. The gluinos subsequently decay  with the leading decay mode being $\tilde g\to q\bar q \tilde \chi_1^0$.}
\label{tab3}
\end{table}

\begin{table}[H]
\begin{center}
\begin{tabular}{l|rc|rc}
\hline\hline
Model & Leading SR & $\mathcal{L}$ (fb$^{-1}$)  & Subleading SR & $\mathcal{L}$ (fb$^{-1}$) \\
\hline
(i)   & 2jl &   14 & 2jm &   20 \\
(ii)  & 2jm &   66 & 2jl &   71 \\
(iii) & 2jm &  180 & 2jl &  260 \\
(iv)  & 2jm &  440 & 2jl &  760 \\
(v)   & 2jm &  970 & 2jl & 1900 \\
(vi)  & 2jm & 2000 & 2jl & 4200 \\
(vii) & 2jm & 3400 & 2jl & 7600 \\
\hline
\end{tabular}\end{center}
\caption{Integrated luminosity for SUSY discovery in the leading and subleading signal regions for 
\gco~Models (i)-(vii) of table \ref{tab1} at \lhc2.}
\label{tab4}
\end{table}

\begin{table}[H]
\begin{center}
\begin{tabular}{l|c|ccccccc}
\hline\hline
Model & $\mathcal{L}$ (fb$^{-1}$) & 2jl & 2jm & 2jt & 4jt & 5j & 6jm & 6jt \\
\hline
(i)   &   14 & \bf5.0 & 4.1 & 1.9 & 1.2 & 2.0 & 1.5 & 0.8 \\
(ii)  &   66 & 4.8 & \bf5.0 & 2.5 & 1.8 & 2.0 & 1.1 & 1.0 \\
(iii) &  180 & 4.2 & \bf5.0 & 2.3 & 1.5 & 1.7 & 1.2 & 0.8 \\
(iv)  &  440 & 3.8 & \bf5.0 & 2.7 & 1.4 & 1.4 & 0.5 & 0.3 \\
(v)   &  970 & 3.6 & \bf5.0 & 3.0 & 1.7 & 1.6 & 1.0 & 1.1 \\
(vi)  & 2000 & 3.5 & \bf5.0 & 2.8 & 1.6 & 1.4 & 0.5 & 0.5 \\
(vii) & 3400 & 3.3 & \bf5.0 & 3.1 & 1.9 & 1.1 & 0.5 & 0.4 \\
\hline
\end{tabular}\end{center}
\caption{{
A grid showing the ratio of signal divided by the square root of the background for the seven
signal regions analyzed for each of the Models (i)-(vii) at the minimum $\mathcal{L}$ 
needed for discovery. The signal regions reaching the 5$\sigma$ threshold 
at the minimum $\mathcal{L}$ needed for discovery are indicated in bold.}}
\label{luminosity}
\end{table}

\begin{figure}[H]
	\includegraphics[width=0.33\textwidth]{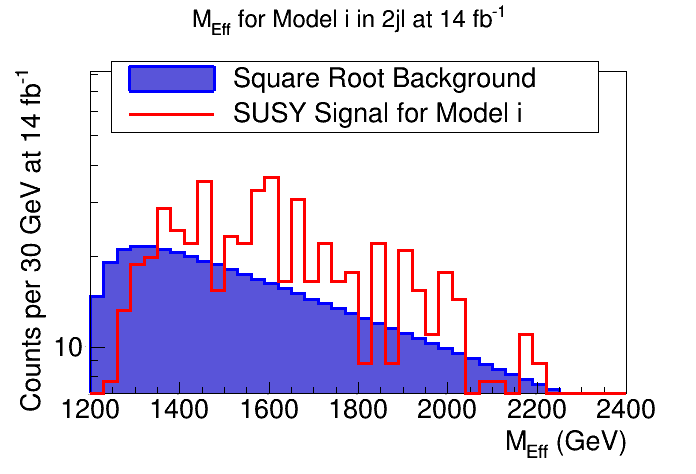}
    \includegraphics[width=0.33\textwidth]{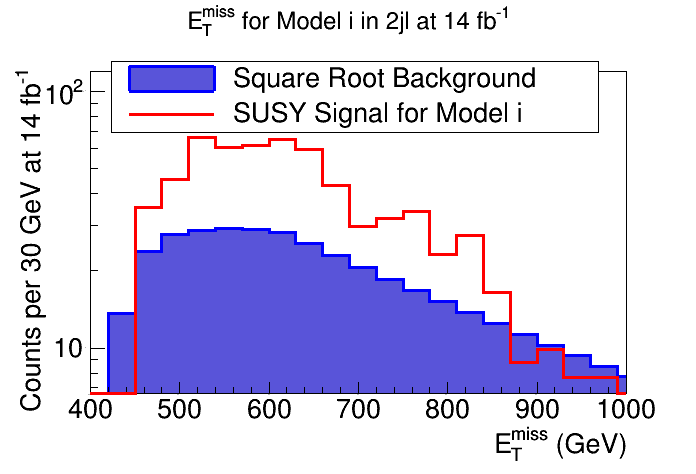}
    \includegraphics[width=0.33\textwidth]{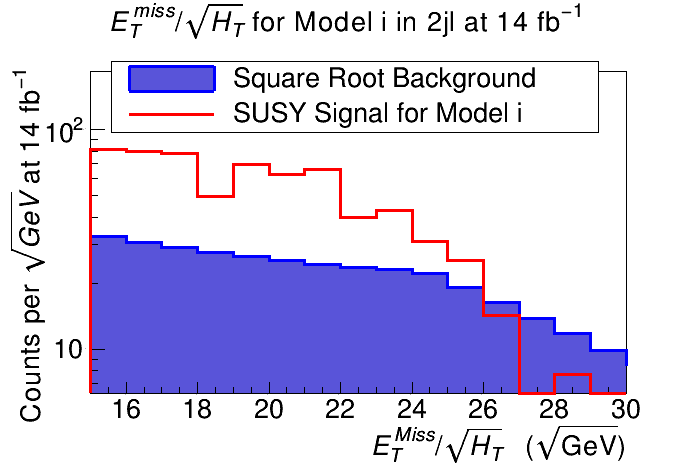}
    \caption{Left panel: Distribution in $M_{\text{eff}}$ for the $2jl$ signal region for \gco~Model~(i). Plotted is the number of counts for the SUSY signal per 30 GeV and the square root of the total SM SNOWMASS backgrounds. The analysis is done at 14 fb$^{-1}$ of integrated luminosity at the energy scale of \lhc2, which gives a $5\sigma$ discovery in this signal region. Middle panel: The same analysis as in the left panel but for $E^{\text{miss}}_T$. Right panel: The same analysis but for $E^{\text{miss}}_T/\sqrt{H_T}$}
    \label{fig-i}
\end{figure} 

\begin{figure}[H]
	\includegraphics[width=0.33\textwidth]{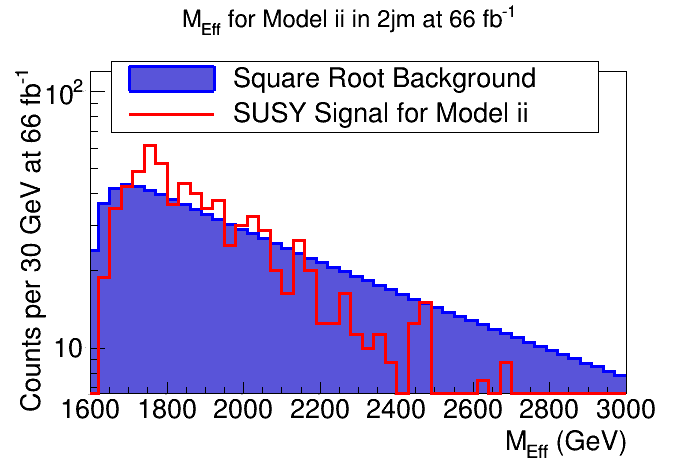}
    \includegraphics[width=0.33\textwidth]{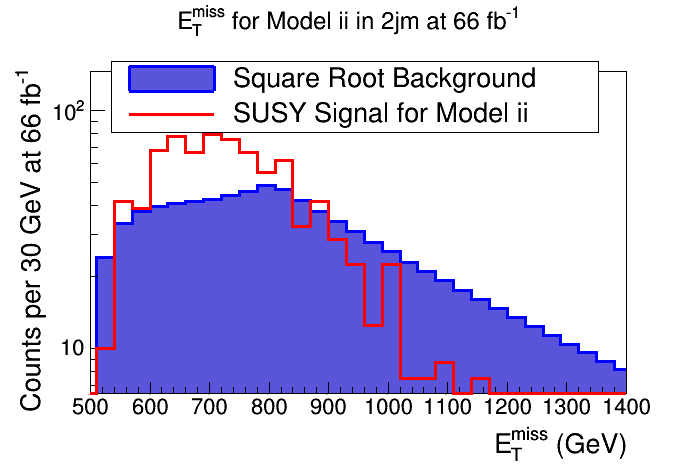}
    \includegraphics[width=0.33\textwidth]{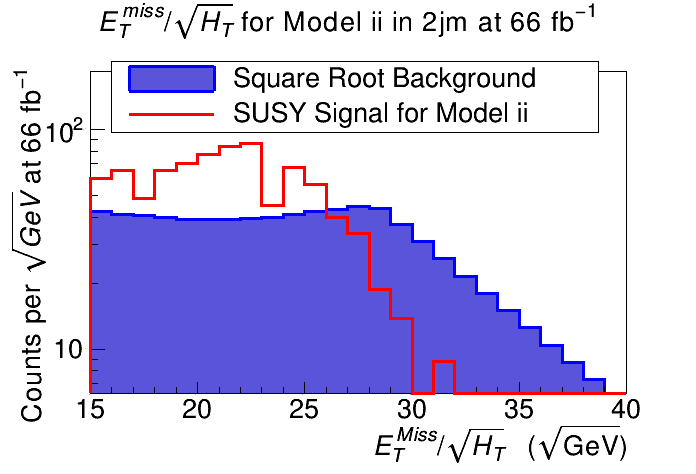}
    \caption{Left panel: Distribution in $M_{\text{eff}}$ for the $2jm$ signal region for \gco~Model~(ii). Plotted is the number of counts for the SUSY signal per 30 GeV and the square root of the total SM SNOWMASS backgrounds. The analysis is done at 66 fb$^{-1}$ of integrated luminosity at the energy scale of \lhc2, which gives a $5\sigma$ discovery in this signal region. Middle panel: The same analysis as in the left panel but for $E^{\text{miss}}_T$.  Right panel: The same analysis but for $E^{\text{miss}}_T/\sqrt{H_T}$}
    \label{fig-ii}
\end{figure} 

\begin{figure}[H]
	\includegraphics[width=0.33\textwidth]{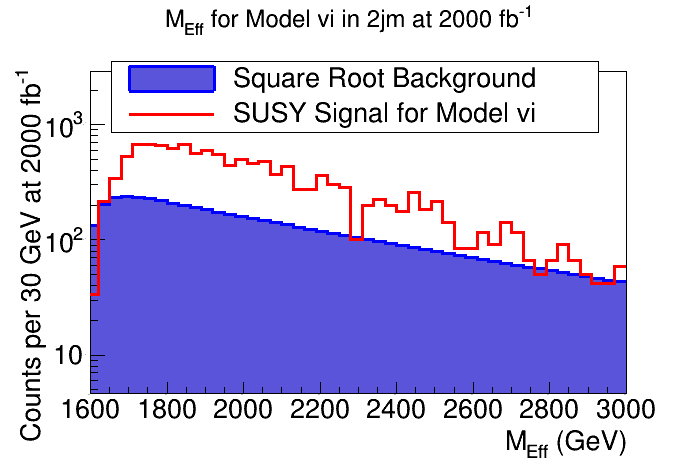}
    \includegraphics[width=0.33\textwidth]{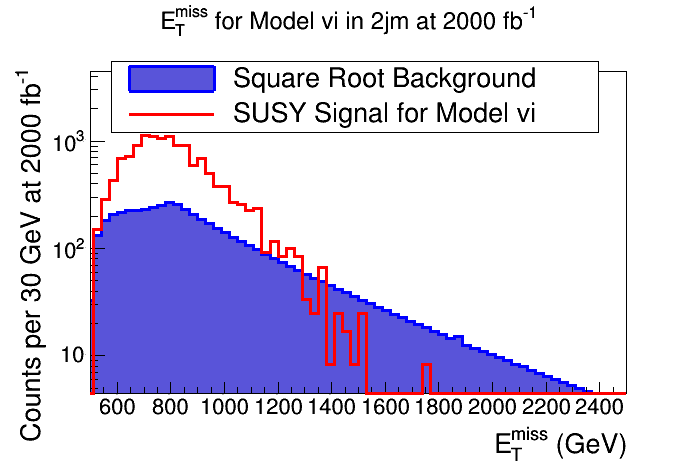}
    \includegraphics[width=0.33\textwidth]{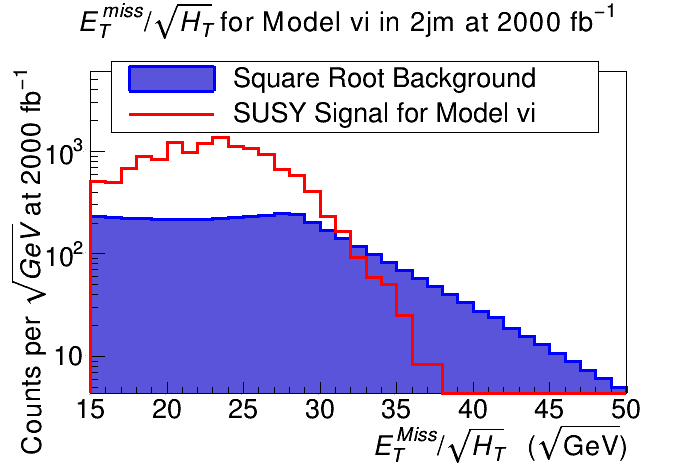}
    \caption{Left panel: Distribution in $M_{\text{eff}}$ for the $2jm$ signal region for \gco~Model~(vi). Plotted is the number of counts for the SUSY signal per 30 GeV and the square root of the total SM SNOWMASS backgrounds. The analysis is done at 2000 fb$^{-1}$ of integrated luminosity at the energy scale of \lhc2, which gives a $5\sigma$ discovery in this signal region. Middle panel: The same analysis as in the left panel but for $E^{\text{miss}}_T$. Right panel: The same analysis but for $E^{\text{miss}}_T/\sqrt{H_T}$}
    \label{fig-vi}
\end{figure} 

\begin{figure}[H]
	\includegraphics[width=0.33\textwidth]{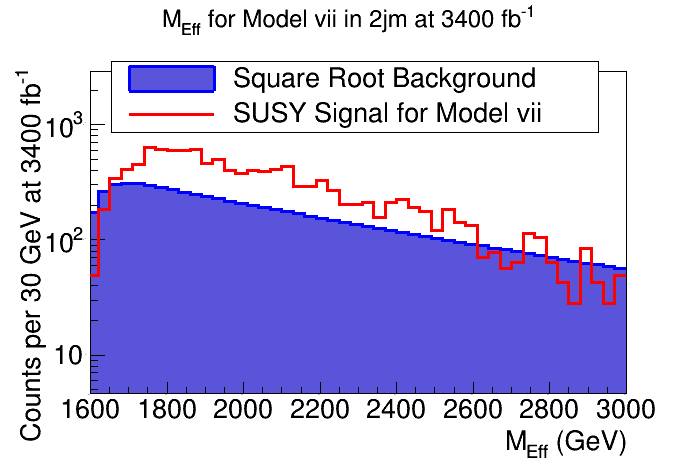}
    \includegraphics[width=0.33\textwidth]{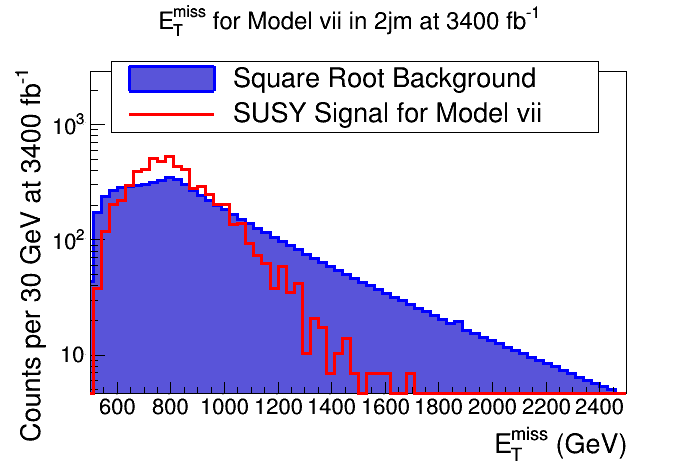}
    \includegraphics[width=0.33\textwidth]{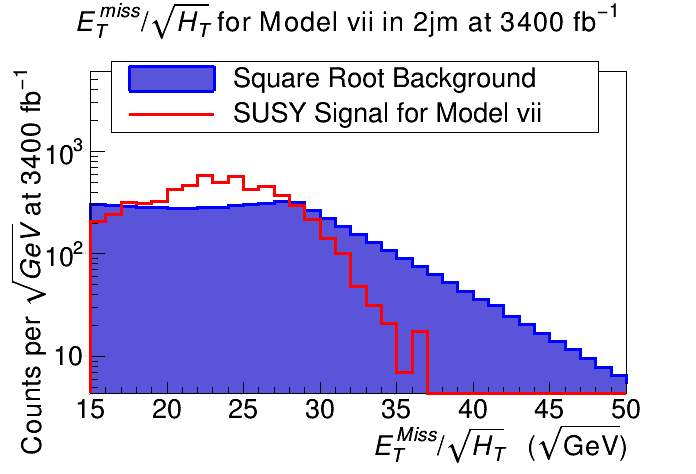}
    \caption{Left panel: Distribution in $M_{\text{eff}}$ for the $2jm$ signal region for \gco~Model~(vii). Plotted is the number of counts for the SUSY signal per 30 GeV and the square root of the total SM SNOWMASS backgrounds. The analysis is done at 3400 fb$^{-1}$ of integrated luminosity at the energy scale of \lhc2, which gives a $5\sigma$ discovery in this signal region. Middle panel: The same analysis as in the left panel but for $E^{\text{miss}}_T$. Left panel: The same analysis but for $E^{\text{miss}}_T/\sqrt{H_T}$}
    \label{fig-vii}
\end{figure}

\subsection{Discussion of results}
In table \ref{tab3} we give an analysis of the SUSY production cross sections, which for the \gco~models considered here is essentially limited to gluino pair production.  The main mechanisms for gluino production are gluon fusion $gg\to  \tilde g \tilde g$ and $q\bar q\to  \tilde g \tilde g$.
Because of the small mass gap between the gluino and neutralino masses needed for \gco, the decay of the gluino is dominated by three-body decay involving a neutralino and quarks, i.e. by the process $\tilde g\to q\bar q \tilde \chi_1^0$. The subleading decay is $\tilde g\to g \tilde \chi_1^0$ which is typically only a few percent with the decay mode $\tilde g\to q\bar q \tilde \chi_1^0$ being as {{as much as} 95\% or more over most of the parameter space, as can be seen in table \ref{tab3} by adding the two columns on the right.
 
{
Using the signal regions defined in~\cite{ATLAS:2015}, it is found  (see table \ref{tab4} and table \ref{luminosity}) that the two leading signal regions for the detection of the decay of the gluino for the \gco~models are $2jl$ and $2jm$ over the entire range of gluino masses given in table \ref{tab1}.  Specifically, in table \ref{luminosity}  we present a grid where the ratio ${\rm signal}/\sqrt{\rm background}$ for each of the Model (i) - (vii) for the seven signal regions is exhibited for an integrated luminosity sufficient for the discovery of the corresponding model. It is seen that for Model (i) $2jl$ is the discovery signal region while for Models (ii)-(vii), $2jm$ is the signal region for discovery. All other signal regions lie significantly below {{the threshold}. Since for each signal region  up to 12 kinematical variables with cuts are utilized, it is of interest to analyze in further detail as to the most sensitive of these kinematical quantities  that allow us 
 to discriminate the signal over the background. It is found that among the twelve kinematical variables, three stand out as the most sensitive for our analysis. These are $M_{eff}$, $E_T^{\rm miss}$, and $E_T^{\rm miss}/\sqrt{H_T}$. } We illustrate this for the cases (i), (ii), (vi), and (vii) below.

For Model (i) the sparticle mass spectrum is exhibited in fig.~\ref{fig2}, which shows the mass hierarchy of all the sparticle states. The sparticles lying in the mass region below $2.5$ TeV have the mass hierarchy 
\begin{align}
 m_{\tilde \chi_1^0}  <  m_{\tilde g} <  m_{\tilde \chi_1^{\pm}}\simeq  m_{\tilde \chi_2^0} 
 <  m_{\tilde t_1}  <  m_{\tilde \chi_2^{\pm}}\simeq  m_{\tilde \chi_{3,4}^0}  \,.
 \end{align}
 In the sparticle landscape this is the mass hierarchy labeled NUSP14 in the nomenclature of 
\cite{Feldman:2009zc}. This hierarchy is a useful guide to what one may expect in this class of models. In the left panel of fig.~\ref{fig-i} we exhibit the 2jl signal region for this model, where we plot the number of SUSY signal events (red) and the square root of the total standard model background (blue) vs. $M_{Eff}$.  The analysis is done for \lhc2~at an integrated luminosity of 14 fb$^{-1}$. {Similar analyses are given in the middle  panel of fig.~\ref{fig-i} for $E_T^{miss}$  and in the 
right panel of   fig.~\ref{fig-i} for $E_T^{miss}/\sqrt{H_T}$.}
Very similar analyses were carried out  for Models (ii), (vi), and (vii) in fig.~\ref{fig-ii}, fig.~\ref{fig-vi}, and
fig.~\ref{fig-vii}. 
Here, however, the leading signal region is 2jm and the analyses are done at an integrated luminosity of 66 fb${-1}$ for Model (ii) in fig.~\ref{fig-ii},
at 2000 fb$^{-1}$ for Model (vi) in fig.~\ref{fig-vi}, and at 3400 fb$^{-1}$ for Model (vii) in fig.~\ref{fig-vii}.
 In each of these cases the SUSY signal meets the $5\sigma$ limit needed for discovery.  A tabulation of the integrated luminosity needed for the discovery of  all the models of table \ref{tab1} is given in table \ref{tab4}. 

The analysis of fig.~\ref{fig-i},  fig.~\ref{fig-ii}, fig.~\ref{fig-vi}, fig.~\ref{fig-vii} and of table \ref{tab4} shows the remarkable reduction in the 
potential of \lhc2 for the discovery of gluino if the gluino mass lies in \gco~region. 
{A further illustration of this result is  given}
in fig.~\ref{fig:lumin} where the largest gluino mass discoverable as a function of the integrated luminosity
at \lhc2 is given if the gluino mass lies in the \gco~region. Specifically one finds that even with $\sim 3000$ fb$^{-1}$ 
of integrated luminosity \lhc2 will probe a gluino mass of $\sim 1200$ GeV in this case. In contrast away from the
\gco~region, the discovery potential of \lhc2 increases in a dramatic fashion as illustrated by the red dot 
in the upper left hand corner of fig.~\ref{fig:lumin}, which is taken from the ATLAS analysis ~\cite{ATLAS:2015} using $3.2$ fb$^{-1}$ of integrated luminosity at \lhc2. Thus the analysis of fig.~\ref{fig:lumin} shows that if the gluino mass lies
in the \gco~region, its mass could be much smaller than the currents lower limits, where the mass gap between
the gluino mass and the neutralino mass is large.  

\begin{figure}[H]
\begin{center}
	\includegraphics[width=0.6\textwidth]{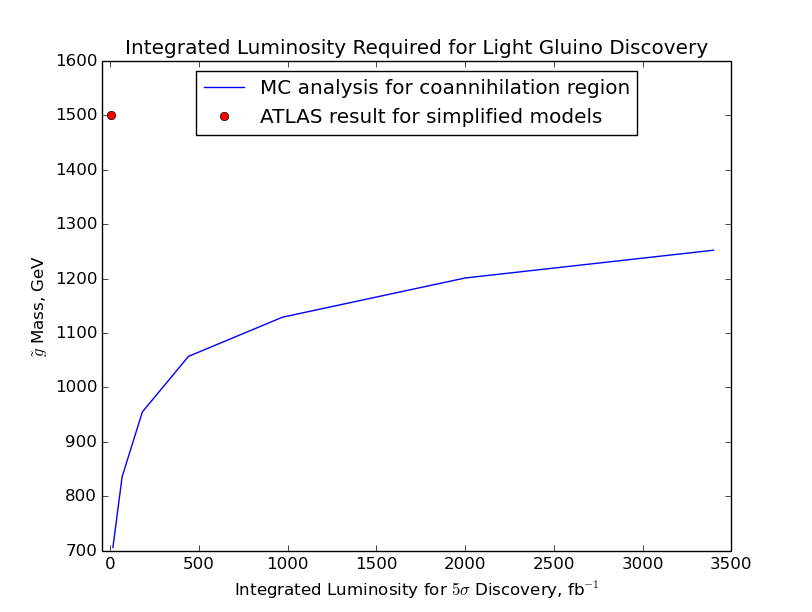}
	    \caption{{Integrated luminosity required at \lhc2  for 5$\sigma$ discovery of  gluino in the \gco~region as a function of the gluino mass. For comparison ATLAS result from~\cite{ATLAS:2015}
	    using simplified models is also exhibited. One finds that the exclusion limit for the gluino mass
	    in the coannihilation region is much lower than the ATLAS result.}}
    \label{fig:lumin}
\end{center}
\end{figure} 

\begin{figure}[H]
\begin{center}
	\includegraphics[width=0.6\textwidth]{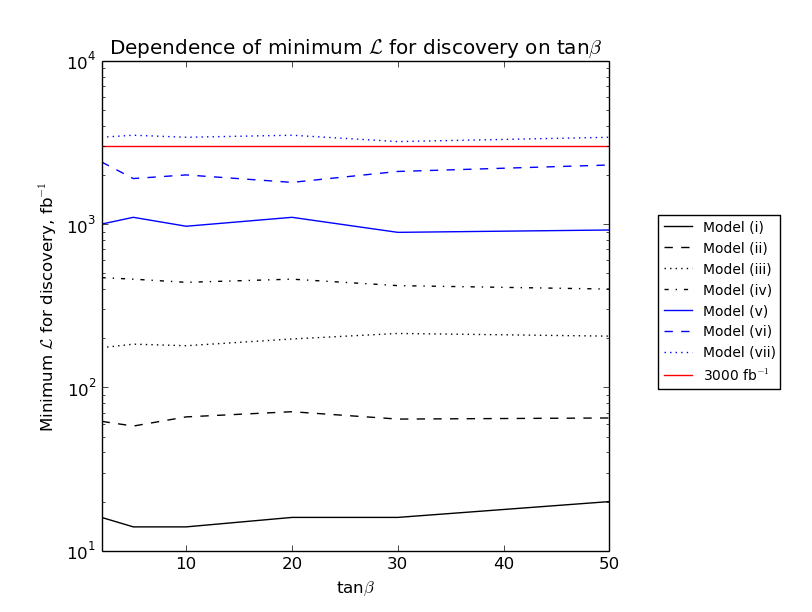}
	    \caption{{Integrated luminosity required at \lhc2  for 5$\sigma$ discovery of a gluino in the	 	     \gco~as a function of $\tan\beta$ for each of the Models (i)-(vii). The dependence
	    of the integrated luminosity required for 5$\sigma$ discovery is seen to have only a mild
	    dependence on $\tan\beta$.}	    
	    }
    \label{fig:tanBeta-lumin}
\end{center}
\end{figure} 

\subsection{Signal Region Optimization\label{optimization}}
{{
Because the signal regions as defined in~\cite{ATLAS:2015} are for generic simplified model light gluinos, it is advantageous to consider whether the signals can be improved upon by altering cuts to one or several of the variables in table~\ref{tab4}. It was found that very modest reduction (between 2 and 10 percent) in integrated luminosity required for 5$\sigma$ discovery can be achieved in the $2jm$ signal region by altering the cut on the variable $E_T^{\rm miss}/\sqrt{H_T}$.  To demonstrate this choice, figure~\ref{noHT} shows the results of the $2jm$ signal region in Models (i), (ii), (vi), and (vii) prior to any cuts on $E_T^{\rm miss}/\sqrt{H_T}$, at the required integrated luminosity for discovery for that model.

\begin{figure}[H]
	\includegraphics[width=0.5\textwidth]{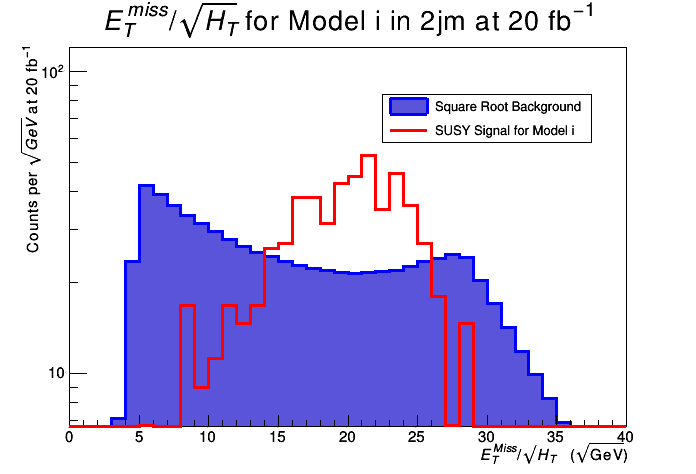}
    \includegraphics[width=0.5\textwidth]{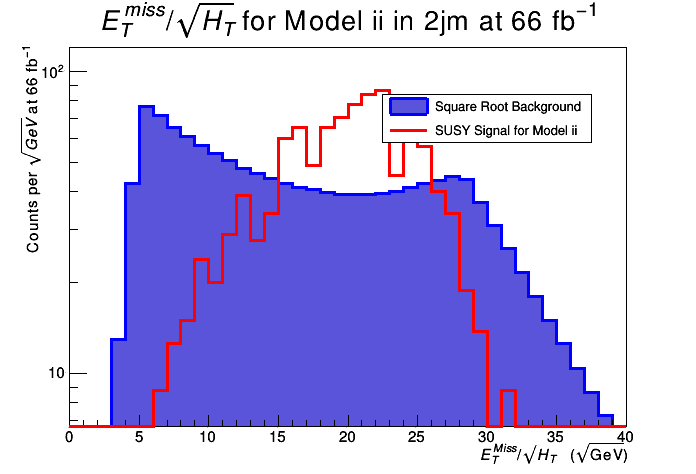}
    \includegraphics[width=0.5\textwidth]{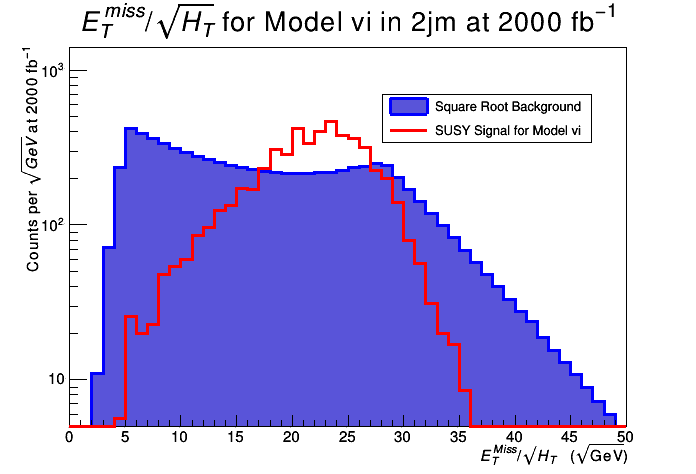}
    \includegraphics[width=0.5\textwidth]{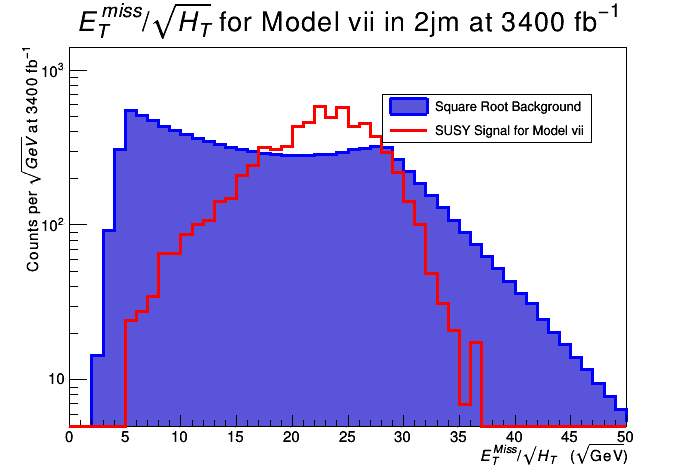}
    \caption{Upper Left panel: Distribution in $E_T^{\rm miss}/\sqrt{H_T}$ for the $2jm$ signal region for \gco~Model~(i) prior to any cuts in $E_T^{\rm miss}/\sqrt{H_T}$. The analysis is done at 20 fb$^{-1}$ of integrated luminosity at the energy scale of \lhc2, which gives a $5\sigma$ discovery in this signal region. Upper Right Panel: The same analysis but for Model~(ii) at 66 fb$^{-1}$ of integrated luminosity.  Lower Left Panel: The same analysis but for Model~(vi) at 2000 fb$^{-1}$ of integrated luminosity. Lower Right Panel: The same analysis but for Model~(vii) at 3400 fb$^{-1}$ of integrated luminosity.}
    \label{noHT}
\end{figure} 

By analyzing these figures, the original choice to require $E_T^{\rm miss}/\sqrt{H_T}>15\sqrt{\text{GeV}}$ (see table~\ref{tab:SR}) is shown to be a very good choice, as this is the point where the SUSY signal (in red) overtakes the square root of the SM background (in blue). However, some improvement can be made by also requiring $E_T^{\rm miss}/\sqrt{H_T}$ to be less than a specific critical value, where the SUSY signal drops below the square root of the SM background.  This occurs around the value of $25~\sqrt{\text{GeV}}$ for the lighter models (Model (i) and (ii)) and around $30~\sqrt{\text{GeV}}$ for the heavier models (Model (vi) and (vii)).  For the optimization, a value of $30~\sqrt{\text{GeV}}$ was chosen to optimize the fit for the heaviest gluino models, thereby extending the reach of the LHC. We call this signal region $2jm$-$HT$. Table~\ref{tab:opt} indicates the improvement by modifying the cut on $E_T^{\rm miss}/\sqrt{H_T}$ in this way, while figure~\ref{30HT} displays the binned count data after the optimized cut. {Further improvement may be possible by investigating the substructure of the jets themselves, as discussed in~\cite{Han:2015lha}.} 

\begin{figure}[H]
	\includegraphics[width=0.5\textwidth]{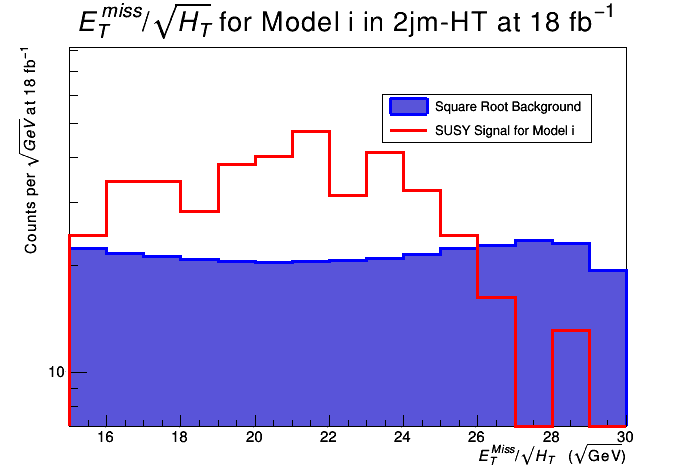}
    \includegraphics[width=0.5\textwidth]{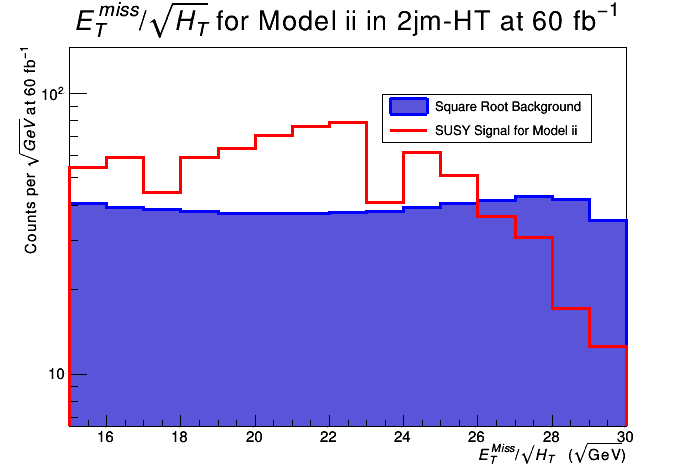}
    \includegraphics[width=0.5\textwidth]{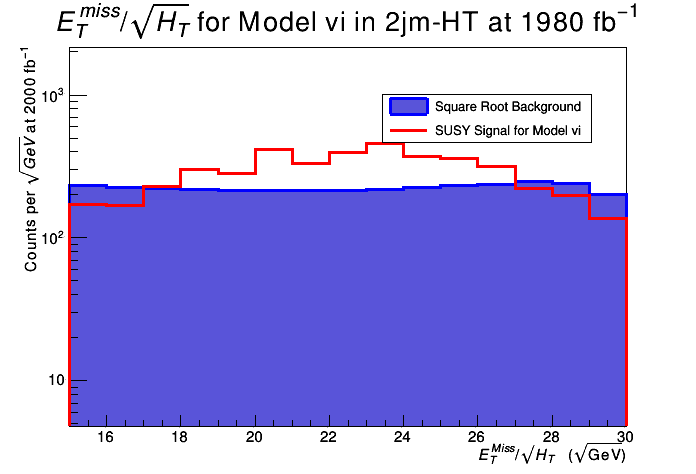}
    \includegraphics[width=0.5\textwidth]{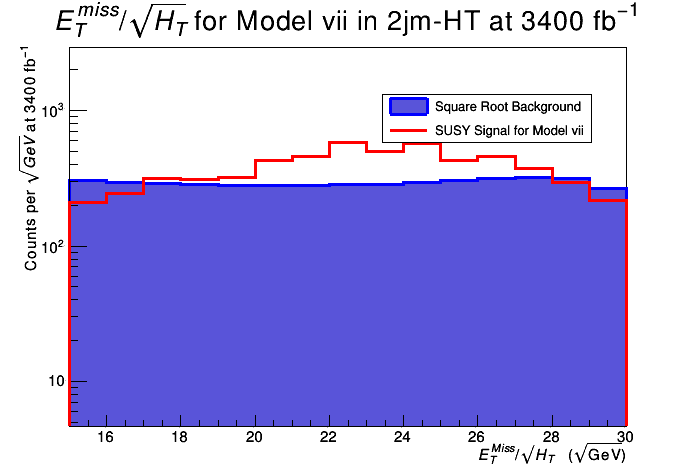}
    \caption{Upper Left panel: Distribution in $E_T^{\rm miss}/\sqrt{H_T}$ for the $2jm-HT$ signal region for \gco~Model~(i) after optimized cuts in $E_T^{\rm miss}/\sqrt{H_T}$. The analysis is done at 18 fb$^{-1}$ of integrated luminosity at the energy scale of \lhc2, which gives a $5\sigma$ discovery in this signal region. Upper Right Panel: The same analysis but for Model~(ii) at 60 fb$^{-1}$ of integrated luminosity.  Lower Left Panel: The same analysis but for Model~(vi) at 1980 fb$^{-1}$ of integrated luminosity. Lower Right Panel: The same analysis but for Model~(vii) at 3400 fb$^{-1}$ of integrated luminosity.}
    \label{30HT}
\end{figure} 
}

\begin{table}[H]
\begin{center}
\begin{tabular}{l|cc}
\hline\hline
Model & $\mathcal{L}$ (fb$^{-1}$) in 2jm  & $\mathcal{L}$ (fb$^{-1}$) in 2jm-HT \\
\hline
(i)   & 20 & 18 \\
(ii)  & 66 & 60 \\
(iii) & 180 & 170 \\
(iv)  & 440 & 420 \\
(v)   & 970 & 950 \\
(vi)  & 2000 & 1980 \\
(vii) & 3400 & 3400 \\
\hline
\end{tabular}\end{center}
\caption{Integrated luminosity for SUSY discovery in $2jm$ and $2jm$-$HT$, where $2jm$-$HT$ requires that $15<E_T^{\rm miss}/\sqrt{H_T}<30$ in units of $\sqrt{\text{GeV}}$}
\label{tab:opt}
\end{table}

\section{Gluino Coannihilation, LHC and Direct Detection of CDM}\label{sec:dm}

We discuss now further details of the \gco~model. 
For the case when the gluino is the NLSP the processes that enter in coannihilation are

\begin{align}
\tilde \chi^0\tilde \chi^0 \to SM\,,
~\tilde \chi^0\tilde g \to SM'\,,
~\tilde g \tilde g \to SM''\,,
\end{align}
where $SM, SM', SM''$ stand for Standard Model states. In the analysis of coannihilation the 
mass gap between the LSP and the NLSP plays a critical role. Thus in general for 
processes  $\tilde\chi_i \tilde \chi_j \to SM$ the coannihilation is controlled by the ratio
$\delta_i$ \cite{Griest:1990kh,Gondolo:1990dk,Nath:1992ty}

 \beqn \delta_i=\frac{n_i^{\rm eq}}{n^{\rm eq}} =
\frac{g_i(1+\Delta_i)^{3/2} e^{-\Delta_i x}} {\sum_j g_j
(1+\Delta_j)^{3/2}e^{-\Delta_j x}}\ , 
\label{coann.1}
\eeqn 
where $g_i$ are the
degrees of freedom of $\chi_i$, $x={m_1}/{T}$
and $\Delta_i =(m_i-m_1)/m_1$, with $m_1$ defined as the 
LSP mass and where   $e^{-\Delta_i x}$ is the Boltzmann suppression 
factor. The relic density involves the integral 

\beqn
J_{x_f}=\int_{x_f}^{\infty} x^{-2} \langle \sigma_{\rm eff} v \rangle dx~,
\label{relic1}
\eeqn
where $v$ is the relative velocity of annihilating  supersymmetric particles,
$\langle \sigma_{\rm eff}v \rangle$ is the thermally averaged cross section times the relative velocity, and $x_f$ is the freeze out temperature.  
The 
$\sigma_{\rm eff}$ that enters  Eq.(\ref{relic1}) has the  approximate form
\beqn
\sigma_{\rm eff}\simeq \sigma_{\tilde g \tilde g} \delta^2_{\na}\left(\delta^2 + 2\delta \frac{\sigma_{\na\tilde g}}{\sigma_{\tilde g \tilde g}} + 
\frac{\sigma_{\na\na}}{\sigma_{\tilde g \tilde g} }\right)~,
\label{relic2}
\eeqn
Here $\delta =\delta_{\tilde g}/ \delta_{\na}$ and  $\delta_i$ are defined by Eq.(\ref{coann.1}).
Numerical analysis shows that  $\sigma_{eff}$ is dominated by the  processes involving the gluino
{{such that a} smaller the mass gap between the gluino and the LSP {{leads to more dominant} gluino processes.
The relic density depends critically on the mass gap and 
 we wish to keep  the relic density constant as the gluino mass increases. However, an increasing 
gluino mass reduces the cross section for the annihilating gluino. In oder to compensate for the 
reduction in the gluino annihilation cross section so that $\sigma_{\rm eff}$  remains constant, the mass gap
between the gluino and the LSP must decrease
(see the first paper of ~\cite{Profumo} and ~\cite{Feldman:2009zc}).
 Specifically one requires that 
\begin{align}
\Delta_{\tilde g\tilde \chi^0}=(m_{\tilde g}/m_{\tilde\chi^0}-1)
\end{align}
must decrease as the LSP mass increases to achieve constancy of the relic density. This is what is seen 
in the analysis of table \ref{tab1}. Further, from the trend in table \ref{tab1} one expects that as the gluino mass gets large 
$m_{\tilde g}/m_{\tilde\chi^0}\to 1$ and $\Delta_{\tilde g\tilde \chi^0} \to 0$ in order to achieve 
 relic density in the desired range. A similar phenomenon was seen in the analysis of~\cite{Kaufman:2015nda}.\\

It is pertinent to ask what the effect of non-perturbative corrections  to  annihilation cross section implies 
regarding the  relic density analysis. In general, non-perturbative corrections 
can be significant near the threshold since here multiple gluon exchanges can occur, which produce the
Sommerfeld enhancement factor.  These effects 
may be approximated by the function  $\cal E$,~\cite{Baer:1998pg} where  
\beqn
{\cal E}_{j}={C_{j} \pi \alpha_s\over \beta}
\left[1-\exp\left\{-{C_{j} \pi \alpha_s\over \beta}\right\}\right]^{-1}\,.
\eeqn
Here $C_{j=g(q)}=1/2(3/2)$ for $\g \g \to gg$ ($\g\g \to q\bar q$),
 where  $\beta = \sqrt{1-4 m^2_{\g}/s}$.
  In \cite{Feldman:2009zc} an analysis was carried out for the relic density including the effect of the Sommerfeld enhancement
 and results compared to the one using 
 \code{micrOMEGAs},  which uses only  perturbative cross section.
  The  analysis of \cite{Feldman:2009zc} shows that the effect of Sommerfeld enhancement is equivalent to an upward shift of
   the gluino mass  by 3-6 GeV without the inclusion of the Sommerfeld enhancement. 
Thus, based on the analysis of \cite{Feldman:2009zc}, inclusion of Sommerfeld enhancement
   would modify the results by only a few percent and our conclusions are not affected in any significant way.

{In addition to the Sommerfeld enhancement  there are also higher order QCD corrections beyond the tree-level prediction given by the code. While the inclusion of the higher order effects is beyond the scope of this
work we can estimate the possible impact of such effects by enlarging the error corridor of the relic density
and determine its impact on the discovery potential of \lhc2 in this case for a given model point.}{Thus in table~\ref{tab:omega-perturb}, the $m_1=m_2$ parameter of Model (i) is adjusted to achieve the 
range of $\Omega h^2$ at the 95\% confidence interval as given in~\cite{Herrmann:2007zr}.

\begin{table}[H]
\begin{center}
\begin{tabular}{l|cccc|cc}
\hline\hline
Model & $m_1=m_2$ & Gluino & Neutralino & $\Omega_{\text{LSP}}h^2$ & Leading SR & $\mathcal{L}$ (fb$^{-1}$)\\
\hline
(i)   & 1400 & 706 & 634 & 0.122 & 2jl & 14\\
(i-a) & 1396 & 706 & 632 & 0.137 & 2jl & 16\\
(i-b) & 1402 & 706 & 635 & 0.093 & 2jl & 15\\
\hline
\end{tabular}\end{center}
\caption{{Effect on Model (i) of perturbing inputs to achieve $\Omega h^2$  encompassing the 95\% confidence range given in~\cite{Herrmann:2007zr}.   The change in the integrated
luminosity needed to discover model (i) with an enlarged range of the relic density is given in the last column 
and the effect is less than 15\%  exhibiting that the analysis is robust. All mass parameters are given in GeV. }}
\label{tab:omega-perturb}
\end{table}

}
}

The connection between LHC physics and dark matter has been discussed in the literature for some time
\cite{lhc-dark} (for a review see~\cite{Nath:2010zj}).
In the context of the \gco~models the connection is even stronger. This is due to the close proximity of the neutralino mass to the gluino mass in this class of  models. Thus a determination of gluino mass at the LHC would indirectly imply a determination of the neutralino mass since it lies within 10\%  of the gluino mass. One can thus make more definitive predictions for the direct detection of dark matter in this case. In general the neutralino has eigencomponents so that $\tilde \chi_1^0= \alpha \tilde b + \beta \tilde w + \gamma \tilde h_1 + \delta \tilde h_2$, where $\tilde b$ is the bino, $\tilde w$ is the wino, and $\tilde h_{1,2}$ are the two Higgsinos of MSSM. The cross section for the direct detection of dark matter depends on the Higgsino content of the neutralino.
{The gaugino and Higgsino eigencomponents for Models (i)-(vii) are given in table \ref{neutralino-components}. 
One may define the Higgsino content of the neutralino by the quantity $\sqrt{\gamma^2 + \delta^2}$. 
From table \ref{neutralino-components} we see that the Higgsino content of the neutralino is typically small 
$<0.05$. A small Higgsino content indicates that the neutralino-proton cross sections will be small in the 
\gco~region.  This is discussed below.}

\begin{table}[H]
\begin{center}
\begin{tabular}{l|rccc}
\hline\hline
Model & $\Delta\times10^3$ & $\beta$ & $\gamma$ & $\delta$ \\
\hline
(i)   & $<1$~~ & -0.001 & 0.021 & -0.008 \\
(ii)  & $<1$~~ & -0.001 & 0.022 & -0.009 \\
(iii) & $<1$~~ & -0.001 & 0.017 & -0.006 \\
(iv)  & 1~~ & -0.003 & 0.042 & -0.027 \\
(v)   & 1~~ & -0.003 & 0.042 & -0.028 \\
(vi)  & $<1$~~ & -0.001 & 0.013 & -0.005 \\
(vii) & 1~~ & -0.002 & 0.037 & -0.024 \\
\hline
\end{tabular}\end{center}
\caption{{Gaugino and Higgsino}
eigencomponents of the neutralino for \gco~Models (i)-(vii), where $\Delta=1-\alpha$.}
\label{neutralino-components}
\end{table}
In table \ref{tab7} we give a computation of the spin-independent and spin-dependent proton-neutralino cross sections. For the spin-independent case one finds $\sigma^{SI}_{p\chi_1^0}$ lying in the range $(1-10)\times 10^{-47}$ cm$^2$.
The next generation LUX-ZEPLIN dark matter experiment~\cite{Cushman:2013zza, Schumann:2015wfa} is projected to reach a sensitivity of $\sim10^{-47}\text{cm}^{-2}$~\cite{Cushman:2013zza,Schumann:2015wfa}. Thus the spin-independent proton-neutralino cross section of the
\gco~models lies largely within the sensitivity range of the next-generation LUX-ZEPLIN experiment.
{A graphical illustration of the spin-independent proton-neutralino cross section is given in figure \ref{fig:dark}. 
Here we are using the Models (i)-(vii) except that we allow $\tan\beta$ to vary between 2-50 and retain only
those model points that satisfy the constraint $\Omega h^2 < 0.123$. 
 One finds that all of the models lie above the neutrino floor, some by 
an order of magnitude or more.  Thus most of the parameter points of figure \ref{fig:dark}  
 appear discoverable by LUX-ZEPLIN.
The neutralino-proton spin dependent cross section $\sigma^{SI}_{p\chi_1^0}$ given by table \ref{tab7} lies in the range $(4-36) \times 10^{-45}$ cm$^{2}$. Here, LUX-ZEPLIN will have a maximum sensitivity of $10^{-42}$ cm$^{2}$, which is about two orders of magnitude smaller in sensitivity than what is needed to test the model in this sector. 

\begin{table}[H]
\begin{center}
\begin{tabulary}{\linewidth}{l|CC|C}
\hline\hline
Model & $R\times\sigma^{SI}_{p\chi_1^0}\times 10^{47}$ & $R\times\sigma^{SD}_{p\chi_1^0}\times 10^{45}$ & Higgsino content \rule{0pt}{2.6ex}\\
\hline
    (i)    & 0.86 & 4.3 &  0.022\\
    (ii)   & 0.92 & 4.9 &  0.024\\
    (iii)  & 0.49 & 1.3 &  0.018\\
    (iv)   & 7.3 & 35 &  0.050\\
    (v)    & 7.2 & 30 &  0.050\\
    (vi)   & 0.29 & 0.50 & 0.014\\
    (vii)  & 5.5 & 19 & 0.044\\
\hline
\end{tabulary}\end{center}
\caption{{ $R\times~\sigma^{SI}_{p\tilde{\chi}_1^0}$ and  $R\times~\sigma^{SD}_{p\tilde{\chi}_1^0}$ in units of cm$^{2}$ for the \gco~models of table \ref{tab1}, where $R=\rho_{\tilde{\chi}_1^0}/\rho_c$ with $\rho_{\tilde{\chi}_1^0}$ the neutralino relic density and $\rho_c$ the critical relic density ($R\sim1$ for the selected \gco~model points). The Higgsino content of the neutralino in each case is also exhibited.}}
\label{tab7}
\end{table}

\begin{figure}[H]
\begin{center}
	\includegraphics[width=0.6\textwidth]{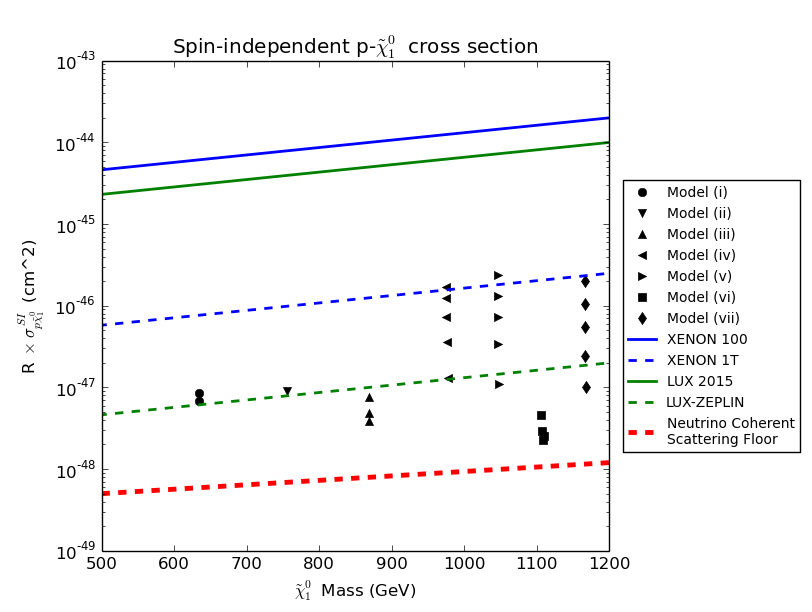}
    \caption{ $R\times~\sigma^{SI}_{p,\tilde{\chi}_1^0}$ as a function of the neutralino mass
   {where the vertical scatter points show the dependence on $\tan\beta$ in the range of $\tan\beta =2-50$
   for a given model.
    Only those parameter points are admitted which satisfy the constraint $\Omega h^2 < 0.123$.}    
    Also {displayed} are the current and projected reaches of the XENON and LUX experiments in the relevant mass range and the neutrino floor~\cite{Akerib:2015rjg,Aprile:2012nq,Ruppin:2014bra,Cushman:2013zza}. }
    \label{fig:dark}
\end{center}
\end{figure}

\section{Conclusion}\label{sec:conclusion}
{

In this work it is shown that in the \gco~region gluino masses as low as 700 GeV consistent with the Higgs boson 
mass constraint and the relic density constraint would have escaped detection at LHC RUN I and \lhc2 with 
3.2 fb$^{-1}$ of integrated luminosity. Our analysis is done within the framework of 
 high-scale supergravity models, allowing for non-universality in the gaugino sector.  An illustrative list of models
 is given in  table \ref{tab1} where the model points all satisfy the Higgs boson mass constraint and the relic density constraint
 consistent with WMAP and Planck.  In the model space analyzed, the gluino is the NLSP, with a mass $\sim1.1$ times the LSP mass. The small mass gap between the gluino and the neutralino is needed 
 to satisfy the relic density constraint via \gco. Requiring that the lightest neutralino supplies all of the thermal relic abundance implies that $\Delta m = m_{\tilde{g}} - m_{\tilde{\chi}_1^0}$ lies in the range 70-100 GeV
over the entire region of the parameter space analyzed in table \ref{tab1}. Because of the small mass gap,  the gluino decay modes such as $\chi_1^{\pm} q_1\bar q_2$ are suppressed and the decay occurs dominantly to $\chi_1^0 q\bar q$
with the subdominant decay mode being $\chi_1^0 g$. 
In the analysis of the Models (i)-(vii) we have used the signal regions used by the ATLAS Collaboration
where an optimization of signal regions was carried out to determine the best regions for
gluino discovery  in the \gco~region.
 It is found that
among the seven signal regions analyzed the leading ones are $2jl$ and $2jm$. It is found that all the models
listed in table \ref{tab1} are discoverable at \lhc2 with up to $\sim 3400$ fb$^{-1}$ of integrated luminosity. }

The implications of the \gco~models for the discovery of dark matter was also discussed. As shown in table \ref{tab7} it is found that the spin-independent neutralino-proton cross section lies in the range $(1-10)\times 10^{-47}$cm$^{-2}$ and this range can be explored in the  next-generation experiments on dark matter, e.g.
 LUX-ZEPLIN~\cite{Cushman:2013zza, Schumann:2015wfa}.  The observed signal from the \gco~region would be a factor of up to ten times stronger than the one from the neutrino floor~\cite{Strigari:2009bq}. Also, as shown in
table \ref{tab7}, the spin-dependent neutralino-proton cross section lies in the range $(0.4-4) \times 10^{-44}$cm$^{2}$. This is about two orders of magnitude smaller than the sensitivity of LUX-ZEPLIN and thus will be more difficult to observe.\\

\textbf{Acknowledgments: }
This research was supported in part by the NSF Grant PHY-1314774.\\

\end{document}